# Ghost spintronic THz-emitter-array microscope

Si-Chao Chen[1,2,†], Zheng Feng[3,†], Jiang Li[1,3,†], Wei Tan[3,†], Liang-Hui Du[1,3], Jianwang Cai[4], Yuncan Ma[1], Kang He[5], Haifeng Ding[5], Zhao-Hui Zhai[1,3], Ze-Ren Li[1], Cheng-Wei Qiu[6], Xi-Cheng Zhang[7], and Li-Guo Zhu[1,3,*]

[1]Institute of Fluid Physics, China Academy of Engineering Physics, Mianyang, Sichuan 621900, China,

[2]Department of Optics and Optical Engineering, University of Science and Technology of China, Hefei, Anhui 230026, China,

[3]Microsystem & Terahertz Research Center, China Academy of Engineering Physics, Chengdu, Sichuan 610200, China,

[4]Beijing National Laboratory for Condensed Matter Physics, Institute of Physics, Chinese Academy of Sciences, Beijing 100190, China,

[5]National Laboratory of Solid-State Microstructures and Department of Physics, Nanjing University, Nanjing, Jiangsu 210093, China,

[6]Department of Electrical and Computer Engineering, National University of Singapore, 4 Engineering Drive 3, Singapore 117583, Singapore,

[7]The Institute of Optics, University of Rochester, Rochester, NY 14627, USA,

†These authors contributed equally to this work.

*email: zhuliguo@tsinghua.org.cn

## Abstract

Terahertz (THz) wave shows great potential in non-destructive testing, bio detection and cancer imaging. Recent progresses on THz wave near-field probes/apertures enable mechanically raster scanning of an object's surface in the near-field region, while an efficient, non-scanning, non-invasive, deeply sub-diffraction-limited imaging still remains challenging. Here, we demonstrate a THz near-field microscopy using a reconfigurable spintronic THz emitter array (STEA) with computational ghost imaging. By illuminating an object with the reconfigurable STEA in near field and computing the correlation measurements, we reconstruct its image with deeply sub-diffraction resolution. By circulating an external magnetic field, the in-line polarization rotation of THz waves is realized, making the fused image contrast polarization-free. The time-of-flight (TOF) measurements of coherent THz pulses further enables to resolve objects at different distances or depths. The demonstrated ghost spintronic THz-emitter-array microscope (GHOSTEAM) is a new imaging tool for THz near-field real-time imaging (with potential of video framerate), especially opening up paradigm-shift opportunities in non-intrusive label-free bioimaging in a broadband frequency range from 0.1 THz to 30 THz (namely 3.3-1000 $cm^{-1}$).

## Introduction

Terahertz waves (0.1—10 THz) with the unique properties[1,2], such as the nonionizing photon energy, spectral fingerprint, and transparency for most nonpolar materials, have attracted many research hotpots and enable a lot of applications, such as non-destructive testing[3], bio detection[4-6] and cancer imaging[7]. However, the long wavelength of THz wave (1 THz ~ 300 μm) confines THz imaging with resolution greater than millimeter scale when merely using conventional imaging methods, due to the well-known Rayleigh diffraction limit, and thus restricts it from many applications such as cellular imaging. Imaging with evanescent wave in near field paves a way towards deep-subwavelength resolution and is desirable particularly for imaging at long wavelength, for example at THz frequencies. The main progress for near-field imaging is efficiently mapping the object-modulated evanescent wave with near-field probes. AFM-, or STM-tip-enhanced THz probes[8,9] or micro-antenna/aperture THz probes[10] achieve micrometer- or even atomic resolution[9], however require to mechanically raster scan the surface of an object pixel by pixel, with relatively poor signal-to-noise ratio (SNR). Near-field ghost imaging techniques[11-14] have been experimentally demonstrated to increase SNR by $\sqrt{N}$ times (N denotes pixel number of a digital picture) over raster scanning[15]. In these approaches, THz images are spatially encoded in near field with deterministic patterns (e.g. Walsh-Hadamard matrix[11-16]), total intensities (or electric field amplitudes) of encoded THz images are collected and detected with a single-pixel detector in far field. After postprocessing by using computational algorithms to correlate the detected THz intensities (or electric field amplitudes) with deterministic patterns[17,18], near-field images are reconstructed. In this ghost imaging scheme, the sub-wavelength spatial information "hidden" in the diffracted far field distribution can be recovered from the intensities (or amplitude fields) recorded by a single-pixel detector, while high-performance near-field THz focal-plane-array camera is still lagging behind.

To encode THz images, photogenerated spatial THz wave modulators in semiconductors have been proposed to act as reconfigurable encoding masks[11-13]. However, THz wave amplitude passing through sub-wavelength apertures follows the scaling rule $1/a^3$ (a denotes the diameter of apertures on mask)[19], which fails to image deep-subwavelength structures.

Alternatively, to avoid the scaling rule, directly detecting THz image[14] or generating patterned THz waves[16] by encoded femtosecond (fs) laser pulses in nonlinear electrooptic (EO) crystal (such as ZnTe) have been proposed. However, sub-wavelength structures are rapidly blurred upon propagation in 100s-μm-thick EO crystals[14] (Supplementary Fig. S1). Although time-resolved single-pixel detection of THz pulses has been theoretically proposed to recover higher-resolution image[16], it requires full-wave measurements of THz pulses in time domain and consequently consumes extra huge detection and encounters with inverse problem. Furthermore, it only applies to recovering 2D-structured images. And constrained by the electromagnetic boundary conditions (Supplementary Fig. S2), the distribution of polarized THz field in sub-wavelength structures presents a "distorted" image[11-14], and might lead to a misunderstanding of the object's morphology.

In this work, we utilize spintronic THz emitter (STE) to illuminate an object in near field. Spintronic THz emitter is a novel type of THz emitter based on the spin related effects in ferromagnetic/non-magnetic (FM/NM) heterostructures[20-21], which is typically of few-nanometer thick but offers comparable generation efficiency to 1-mm-thick ZnTe crystal[22-24]. Fascinatingly, it fully covers 0.1-30 THz frequency range[23], which is superior to all the current solid emitters. Up to now, all the applications of STE only focus on its far-field properties. The highly efficient few-nanometer-thick STE enables illuminating an object at extremely near-field, which allows deeply breaking the diffraction limit. The key challenge for using such STE in near-field imaging is how to map the object-modulated THz field without near-field scanning probes. Based on STE, we develop a near-field illuminating "array" (STEA) whose "elements" are coherent and individually programmable by photoexcitation. All the elements are individually reconfigurable in binary with either "on" or "off" states. And further by utilizing the programmable near-field illuminating STEA and far-field single-pixel detection, we designed and demonstrate a novel ghost spintronic THz-emitter-array microscope (GHOSTEAM) for THz wave deep sub-diffraction-limited near-field imaging. A deeply sub-diffraction resolution limited at single pixel of 6.5 μm was demonstrated with a contrast of more than $57\% \pm 21\%$ (>20% required by Rayleigh criterion[14]) observed in a 6-μm metal gap. Besides, polarization effects on sub-diffraction-limited image was eliminated via post processing of two images with mutually orthogonal polarizations. Moreover, TOF microscopic topography was demonstrated with a 3D silica structure.

## Results

**Concept design**

In our design, as shown in Fig. 1a, the STEA under external magnetic field ***B*** generates spatially structured THz pulses under fs-laser-pulse spatial photoexcitation. The STEA, consisted of W(2nm)/Fe(2nm)/Pt(2nm) tri-layered heterostructure on a transparent MgO substrate (Fig. 1b), performs great THz conversion efficiency in terms of output amplitude, which is comparable with 1-mm ZnTe[23]. The STEA is capped with a 150-nm SiO$_2$ layer ($n$ = 1.97) for protecting it from being damaged by the fs laser. The output THz electric field $E(t)$ is linearly polarized and the polarization is perpendicular to the applied magnetic field $B$, as described by $E(t) \propto (J_c = J_s \times B)$[20-24], where $J_s$ represents the spin current induced by the fs laser and $J_c$ represents the charge current converted in the nonmagnetic metals under $B$. To perform ghost imaging (see Methods for details), the Walsh-Hadamard matrix[25] was used to code the STEA due to its best noise suppression performance among various measurement matrices[11-15]. The patterns programmed in order of Walsh-Hadamard masks, were spatially encoded on the excitation fs-laser beam by a digital micromirror device (DMD) with switching time of 5 µs. On the exit surface of the STEA, the profile of the output THz pulse is as accurate as the excitation fs laser, because the 150 nm propagation distance in SiO$_2$ protective layer is too thin for the THz wave to be diffracted (150 nm ≈ 5 × 10$^{-4}$λ$_0$/$n$, where λ$_0$ = 600 µm and $n$ = 1.97, see Supplementary Fig. S1 for the theoretical calculation about the near-field evanescent wave). The spatial near-field structure of the THz pulse from STEA consists of individually controllable real-time programmable "elements" up to 128 × 128 (see Supplementary Fig. S3) with either "on" or "off" binary emission states, as shown in Fig. 1b. The pixel size of each "element" is 6.5 µm × 6.5 µm in our experiments, which may also be zoomed in or out by an optical projection imaging system according to a specific requirement. Transmitting through an object placed in the near-field region ($z$ = 150 nm), the structured THz pulses were collected and focused on to a 1-mm-thick (110) ZnTe crystal for single-pixel coherent detection by EO sampling. The peak electric amplitudes were recorded for reconstructing THz sub-diffraction ghost images.

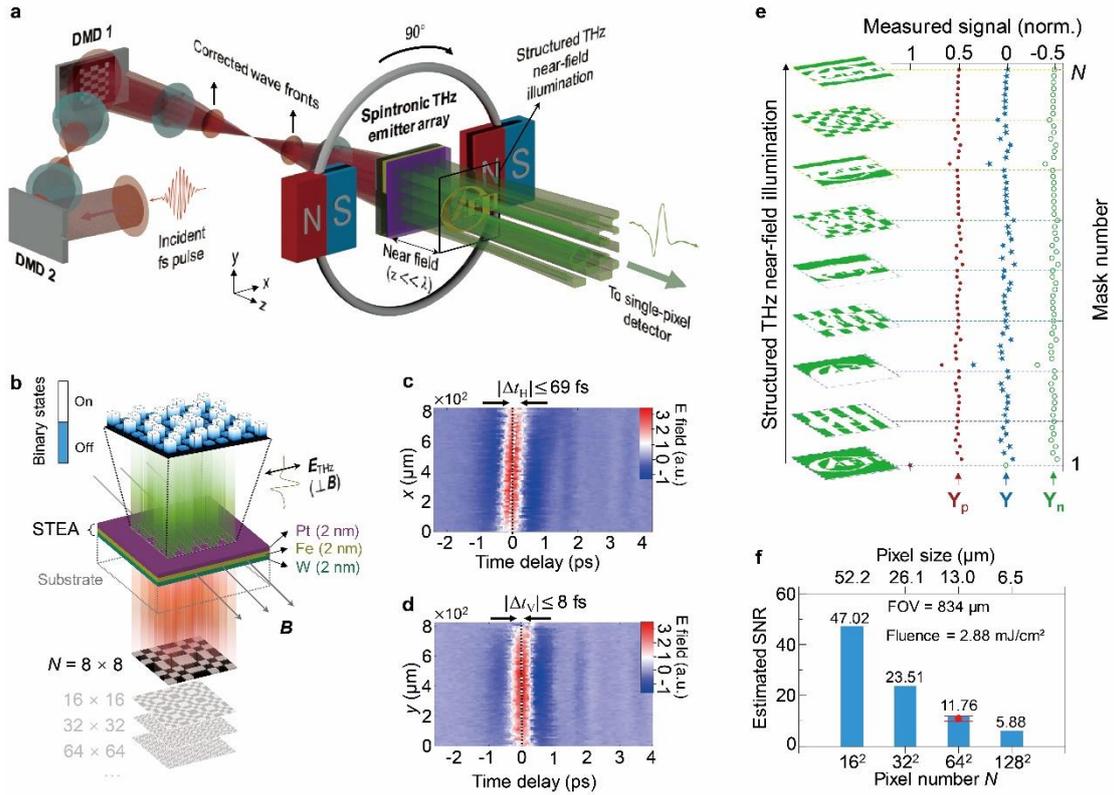

**Figure 1| Schematic of the GHOSTEAM system. a**, Schematic of the GHOSTEAM system. The spintronic THz emitter array (STEA) is excited by two-DMDs-encoded *fs*-laser pulses and generates spatially coded THz pulses. An object "CAEP" was placed in the near-field region ($z\ll\lambda$). The illuminating THz pulse was collected and sent to a single-pixel detector. **b**, Schematic of the STEA, consisting of W(2nm)/Fe(2nm)/Pt(2nm) tri-layered heterostructure and working in binary emission states with polarization perpendicular to applied magnetic field ***B***. **c**,**d**, Spatiotemporal THz waveform along horizontal (**c**) and vertical (**d**) directions. Wavefront indicated by white dotted lines, and time *t* = 0 indicated by black dotted lines. $|\Delta t_H|$, $|\Delta t_V|$ are temporal delays in horizontal and vertical directions, respectively. **e**, Schematic of the detection for ghost imaging. The measured signal **Y** (which is the difference between the positive mask value and negative mask value, **Y** = |**Y**$_p$|-|**Y**$_n$|, in the case of Hadamard multiplexing) from single-pixel detector of an object "CAEP" illuminated by a sequence of prearranged structured THz wave. **f**, Estimated SNRs as a function of pixel number *N* under condition of pump fluence of 2.88 mJ/cm², FOV$_1$ = 834 μm × 834 μm. The red marker with value of 10.92 ± 0.97 is the experimental result (see main text below and Supplementary section 4 for details).

After reflection upon a DMD, the fs-laser pulse suffers from the wavefront tilt in the DMD's deflection direction (horizontal), since the DMD is essentially a reflective blazed grating[26]. The resultant fs-laser pulse with tilted wavefront would badly lose its temporal coherence among individual micromirrors along the deflection direction. To correct the wavefront tilt, another DMD (DMD$_2$) was imaged onto the encoding DMD (DMD$_1$) by a 4$f$ system with 1:1 imaging ratio, as shown in Fig. 1a. Due to the ghost imaging being operated at the fixed time delay (corresponding to the peak of THz pulses), the DMD-induced temporal smearing should be corrected as small as possible, or decoherence (large time delay in wavefront) would result in severely distorted images (see Supplementary Fig. S6). The temporal delay in horizontal |Δ$t_H$| was ultimately corrected less than 69 ± 13 fs (Fig. 1c), which cannot be completely eliminated due to the manufacturing tolerance of two DMDs' tilt angle (see Supplementary Fig. S6). Meanwhile the temporal delay in vertical direction |Δ$t_V$| was measured less than 8 fs (Fig. 1d), which is expectant because the two DMDs only deflect in horizontal direction. The spatiotemporal THz waveforms (Figs. 1c and 1d) were measured by ghost imaging using 64-order Walsh-Hadamard coding 1-D masks (see Methods and Supplementary Fig. S5 for detailed information and see Supplementary GIFs. 1 and 2 for raw data).

An object was illuminated by the coherent programmable STEA with a sequence of prearranged Walsh-Hadamard masks, and the computational-ghost-imaging algorithm was applied to correlate the measured peak amplitude from a single-pixel detector with the sequence of prearranged illuminating THz wave patterns (Fig. 1e, with detailed procedure shown later). The reconfigurable area on STEA, defined as the first imaging field of view FOV$_1$, was measured as 834 μm × 834 μm with coding pixels up to 128 × 128. SNRs of reconstructed images encoded by Walsh-Hadamard masks were numerically determined by a function of pixel number (or pixel size in FOV$_1$, see Methods), which gives a reasonable value of ~47.02 when the FOV$_1$ was coded into 16 × 16 pixels under our experimental condition (pump fluence ~2.88 mJ/cm$^2$, dynamic range of peak field ~1043, pulse fluctuation ~0.7%), as shown in Fig. 1f. Increasing pixel size linearly increases SNR of the reconstructed images. It is worth noting that although the detected THz band was less than 2 THz due to the 90-fs pulse and 3-THz detection bandwidth (1-mm ZnTe), the STE has been demonstrated with fully covering the full THz region, i.e., 0.1—30 THz, if shorter pump pulse (10 fs) and wide-band detector are used[23]. And more experimental details are given in Methods and Supplementary section 2.

**Sub-diffraction ghost imaging**

The STEA were experimentally reconfigured in sequence of 64 × 64-order Walsh-Hadamard matrix to acquire THz sub-diffraction-limited images of an object (see Fig. 2a) positioned at $z$ =150 nm away from the STEA (thickness of the protection $SiO_2$ layer on top of the STEA). Field amplitudes of spatially coded THz waves through the object were measured at the fixed time delay of 0 ps (as indicated by black dotted lines in Figs. 1c and 1d). The reconstructed 64 × 64 ghost images in $FOV_1$ = 834 μm × 834 μm with mutually orthogonal polarizations are shown in Figs. 2b and 2c, whose pixel size are both 13.0 μm × 13.0 μm. And the reconstructed 64 × 64 ghost image in a smaller area $FOV_2$ = 417 μm × 417 μm (indicated by a black dashed box in Fig. 2a) with smaller pixel size of 6.5 μm × 6.5 μm is shown in Fig. 2d. The *y*-dependent THz field distribution across the slit region (indicated by a black dashed arrow in Fig. 2b) was extracted from the associated pixels, as shown in Fig. 2e. Note that every pixel value was averaged from the identical-row pixels within the slit region, with the prior knowledge about the object's horizontal homogeneity in the slit region. A contrast ratio of 57% ± 21% (> 20% required by Rayleigh criterion[12] for distinguishing slits) was observed in the narrowest 6-μm metal slit within $FOV_2$ (see Supplementary section 6 for the quantification process and see Supplementary GIFs. 3-5 for the raw data), which demonstrates that our GHOSTEAM system deeply break the diffraction barrier ~366 μm for the 600-μm THz wave (Rayleigh resolution $0.61\lambda_0$/NA). The spatial resolution of the GHOSTEAM system using this experiment setup is limited at the minimum available pixel size of 6.5 μm, which is dependent on the accuracy of mask patterns on STEA projected by DMD. As the micromirrors of the DMD are arranged in the diamond orientation, the accuracy of coding profile on STEA would decrease when a mask pixel is formed with less micromirrors (see Supplementary Fig. S3). And regardless of diffraction of the coding optical light, the resolution of GHOSTEAM is limited by the propagation distance between the emission surface of STEA to objects (the thickness of the protection 150-nm-thick $SiO_2$ layer in this experiment) which is theoretically estimated as sub-micrometer (see Supplementary Figs. S1).

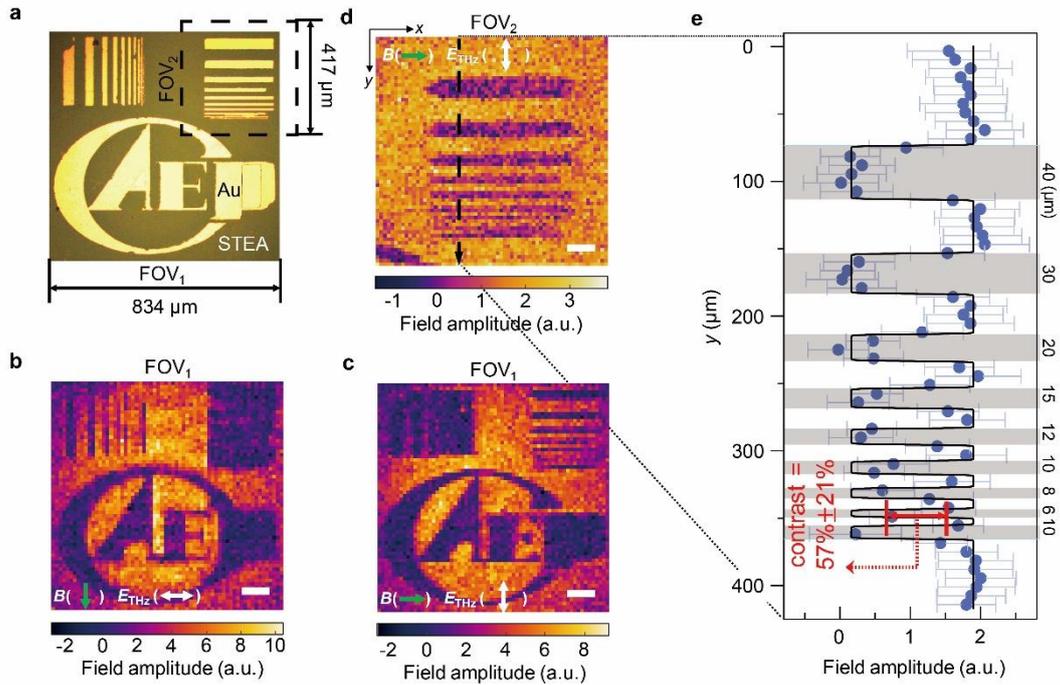

**Figure 2| Sub-diffraction-limited images from GHOSTEAM. a**, Optical image of an object with field of view of $FOV_1$ = 834 μm × 834 μm. The bright regions are gold attached on the 150-nm thick protection $SiO_2$ layer on top of the STEA. THz ghost images in $FOV_1$ with applied magnetic field (green arrows) along vertical direction (**b**) and horizontal direction (**c**). The pixel size and scale bar are 13.0 μm and 100 μm for both images, respectively. **d**, THz ghost image in $FOV_2$ (indicated by the black dashed box in a) with pixel size of 6.5 μm and scale bar of 50 μm. The applied magnetic field ***B*** is along horizontal direction (indicated by the green arrow) and the polarization of THz radiations (indicated by the white double-headed arrow) is perpendicular to ***B***. **e**, Averaged amplitude of THz field along black dashed arrow in b. Blue dots are averaged experimental data and black solid curve is fitted by Boltzmann sigmoidal function (see Supplementary section 5 for details). Gray areas represent the metal regions with corresponding widths indicated. Contrast ratio of 57%±21% at the 6-μm width metal slit.

**Polarization-free image**

Polarization impacts on sub-wavelength imaging (see the dark region where slits along wave polarization in Figs. 2b and 2c), which is expectant due to the electrical field boundary conditions in sub-wavelength scale[11,13,14] (see Supplementary Fig. S2), hinders from accurately resolving the object's morphology. With the advance of in-line rotating THz wave's polarization by an external magnetic field, we secured two images with mutually orthogonal polarizations. This feature allows us to achieve higher-contrast polarization-free image by post-fusing the two images. To do this, 2D Fourier transformation was applied to

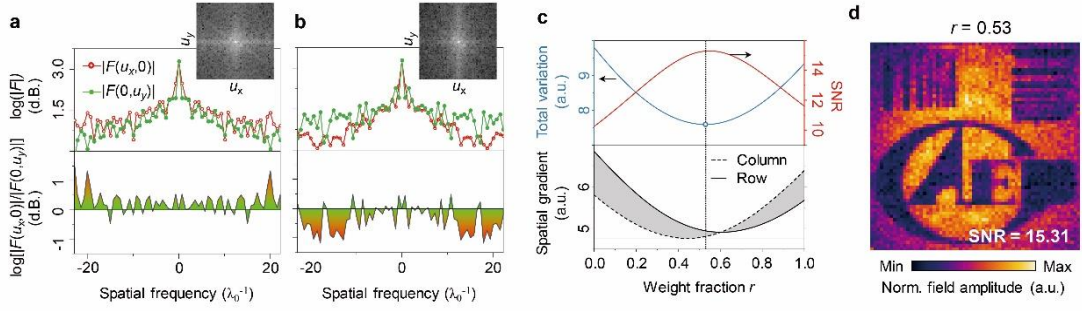

**Figure 3| Polarization-free THz image. a,b**, 2D Fourier transformation of Figs. 2b (**a**) and 2c (**b**). The amplitude $|F(u_x,u_y)|$ ($|F(u_x,0)|$ and $|F(0,u_y)|$ are in red circles and blue dots, respectively) as a function of spatial frequency (upper panels). $\log[|F(u_x,0)|/|F(0,u_y)|]$ is shown in the lower panels for the comparison between horizontal and vertical spatial distribution. **c**, Assessments (Total variation, spatial gradient, see Supplementary section 7 for details) on the fused image as a function of weight fraction *r*. The blue open circle indicates the minimum TV = 7.59 obtained at *r* = 0.53. **d**, Fused polarization-free THz image with optimized SNR = 15.31 at *r* = 0.53.

Figs. 2b and 2c, individually, to quantify the distribution in spatial frequency domain $|F(u_x,u_y)|$ along *x* and *y* directions. The amplitude $|F(u_x,u_y)|$ as functions of spatial frequencies $u_x$ and $u_y$ are shown in Figs. 3a and 3b, which shows loss of high spatial frequencies (> 10/λ) components parallel to $E_{THz}$ in polarization images. To acquire polarization-free THz image $\mathbf{X}_F$, the THz ghost images with horizontal polarization $\mathbf{X}_H$ (Fig. 2b) and with vertical polarization $\mathbf{X}_V$ (Fig. 2c) were fused by the weighted average method, namely $\mathbf{X}_F = r\mathbf{X}_H + (1-r)\mathbf{X}_V$, where *r* represents the weight fraction and belongs to [0,1]. Total variation (TV), which is a common assessment parameter for image's clearness, was chosen to guide the optimization of the fusion process. The TV of $\mathbf{X}_F$ was calculated by

$$\text{TV} = \left(\sum_{i=2}^{64}\sum_{j=2}^{64}\left[\left(\nabla_H \mathbf{X}_F(i,j)\right)^2 + \left(\nabla_V \mathbf{X}_F(i,j)\right)^2\right]\right)^{1/2}, \quad (1)$$

where $\nabla_H$ and $\nabla_V$ are the discretized gradient operators along the horizontal and vertical directions, respectively. At the ratio of *r* = 0.53 in Fig. 3c, the minimal TV was achieved and the corresponding optimal fused image is shown in Fig. 3d. Besides, the SNR (see Supplementary Fig. S13) and directional spatial gradients[26] of the fused image are also given in Fig. 3c. The directional spatial gradients of $\mathbf{X}_F$ were used to quantify the spatial information of the fused image and were calculated by

$$G_{\text{row}} = \left(\sum_{i}^{64}\sum_{j=2}^{64}[(\nabla_H \mathbf{X}_F(i,j))^2]\right)^{1/2}, \quad (2)$$

$$G_{\text{col}} = \left(\sum_{i=2}^{64} \sum_{j}^{64} [(\nabla_V \mathbf{X}_F(i,j))^2]\right)^{1/2}. \tag{3}$$

As shown in Fig. 3c, the fused image at $r$ = 0.53 also shows more homogeneous spatial information in vertical and horizontal directions, as visually presented in Fig. 3d, which overcomes the polarization impacts and is much more similar to the object's morphology.

**Time-of-fight Topography**

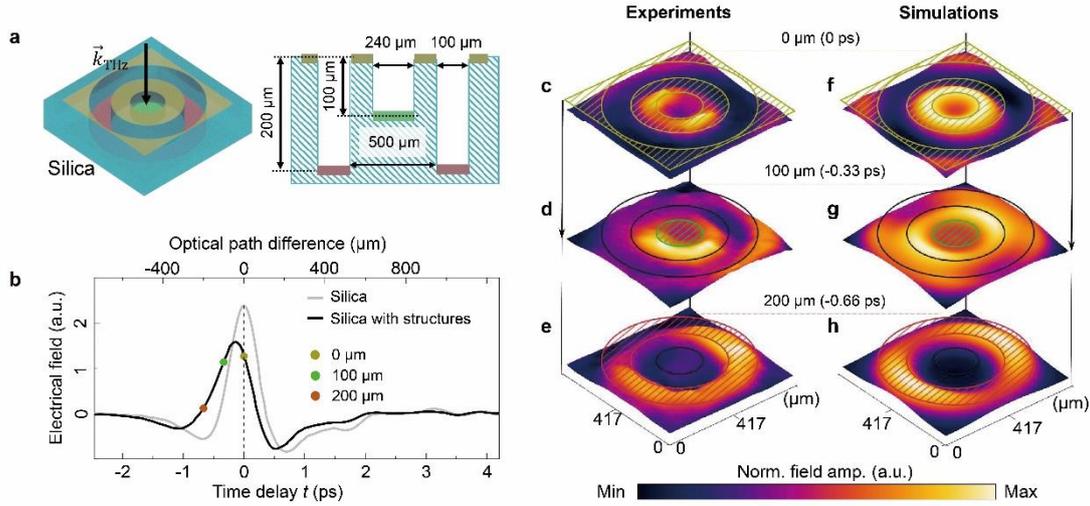

**Figure 4 | Time-of-flight THz microscopic topography using GHOSTEAM. a**, Structure of a prototype 3D object. The object with air grooves on $SiO_2$ has three interfaces [indicated in yellow (upper), green (middle) and red (bottom)]. The arrow indicates the incidence direction of THz pulses with illuminating area of $FOV_1$. **b**, Measured THz waveforms through the sample with and without the structured region. The solid dots indicate EO sampling delay times of TOF with the GHOSTEAM system, corresponding to three interfaces. Experimental sub-diffraction-limited images in height expression for interfaces at 0 μm (**c**), 100 μm (**d**), 200 μm (**e**) respectively. Note that experimental images were denoised using stationary wavelet transformation (see Methods for details). **f-h**, Simulated electrical distributions for three interfaces relevant to c-e, respectively. The color map for each image (c-h) is normalized individually.

Owing to the fact that THz emission from STEA used in our GHOSTEAM system is coherent, we performed TOF measurements[28] on THz pulses from STEA, which enable us to achieve microscopic topography of objects with depth resolution. TOF topography utilizes the optical path differences among different media to acquire the depth ($z$ axis) distribution of an 3D object. An 3D object made of silica by fs-laser manufacturing (see Methods) with three $SiO_2$/air interfaces with air depths of 0 μm, 100 μm and 200 μm

(illustrated in Fig. 4a) was used as a prototype to demonstrate TOF microscopic topography using the GHOSTEAM system. The structured silica was attached on to the external surface of the STEA. Transient THz waveforms from STEA through the object within and without the region of trilayer structure were measured and shown in Fig. 4b. Three THz ghost images relevant to the three interfaces, under the field of view $FOV_1$, were measured at the fixed time delays of 0 ps, -0.33 ps and -0.66 ps (indicated by three solid circles in Fig. 4b), respectively, as shown in Figs. 4c-e. And the reconstructed images agree with the simulated electrical field distributions (Figs. 4f-h, see Methods for simulation details). The sub-wavelength structures (scale of ~$\lambda_0/6$) of the three interfaces are well resolved both experimentally and theoretically (Note that diffraction leads to distortion on reconstruction of the middle interface).

## Discussion

In conclusion, we have presented a novel GHOSTEAM system for THz wave near-field microscopic topography. Compared with other existed THz wave near-field imaging systems, the GHOSTEAM system utilizes a real-time, reconfigurable, coherent STEA for structured near-field illumination. It was numerically and experimentally demonstrated to be able to feature both micrometer-scale-resolved microscope (≤ 6.5 μm) and depth-resolved topography. Future optimizing optical projection imaging system could enable sub-micrometer-scale resolution. Polarization-free sub-diffraction-limited image was also achieved thank to the tunable flexibility of STEA.

The capacity of STE to provide ultra-broadband and ultra-short THz pulses[23] would further enables label-free cellular imaging in 3.3-1000 cm$^{-1}$ wavelength range or ultraprecise topography, a feature lacking in existing THz near-field imaging systems. STE fabricated on flexible substrates[29] would extend our GHOSTEAM system to imaging objects with curved surface. Oscillator *fs*-lasers with MHz repetition rate has been demonstrated to drive STE with comparable THz conversion efficiency to photoconductive antenna[23], which would enable the GHOSTEAM system with much faster acquisition time and consequently real-time near-field imaging with up to ~26 frames/s frame rate (under N = 16 × 16, $SNR_H$ = 10, 50% compression with >95% image fidelity) (see Methods), in contrast to all other existed THz wave near-field ghost imaging[11-14,17] required an amplified fs-laser with few-kHz repetition rate with slow acquisition and complex system. Besides, using shorter fs pulse and wide-band detector, the ultrabroadband property of STE, fully covering 0.1—30 THz frequencies[23], has great potential for realization the broadband THz applications. Additionally, GHOSTEAM system is compatible with compressive sensing (see Supplementary GIFs. 1-5), adaptive sampling[12,30], parallel acquisition in *k*-spac[31], and coherent time-resolved full-wave detection in time domain[17].

## Materials and methods

### Experimental details

THz pulses (0.2-1.7 THz with central frequency of 0.5 THz) were generated in W(2nm)/Fe(2nm)/Pt(2nm) tri-layer heterostructures [placed in a rotatable dc field (|***B***| = 80 mT)] by 800-nm laser (duration of 90 fs, repetition rate of 1 kHz, pump pulse energy of 20 µJ). The THz signals were electro-optic sampled by 1-mm-thick ZnTe (110) in combination with a balanced detector, and were eventually recorded by a lock-in amplifier with integral time of 100 ms (see Supplementary Fig. S4 for details). And all the THz ghost images were multiplexed using Walsh-Hadamard matrix and operated at respective fixed time delays

(e.g., the peak time delays for Figs. 2b-2d and appropriate time delays for Figs. 4c-4e indicated in Fig. 4b). The acquisition time for each mask was 2 s.

To optimize the wavefront of fs pulses, three steps were adopted to determine the positions of $DMD_1$ and $DMD_2$ (Wintech DMD4500, which contains 1140 × 912 diamond-arrayed micromirrors with tilt angle of 12° ± 1° and micromirror pitch of 7.6 μm). First, we acquired the one-to-one image of $DMD_2$ by placing a CCD (See Supplementary Fig. S4c) at the presupposed position of $DMD_1$ to adjust and determine the positions of $DMD_2$ and two lenses (focal lengths of 100 mm). Second, by adjusting the position of $DMD_1$, it can be determined when the images (at the presupposed position of STE) shaped by $DMD_1$ and $DMD_2$ individually were sharp simultaneously. Third, accurately adjust their positions by measuring waveforms of the Walsh-Hadamard masks of #2p and #2n (See Supplementary section 3) until reaching the minimal peak time delay difference between the two waveforms. The tilt angle difference between $DMD_1$ and $DMD_2$ was measured as 0.40° by comparing two DMDs' zero-order diffraction angle in the case of normal incidence.

**Computational ghost imaging**

Let **O** represent the pixelated imaging target, consisting of $N$ unknown elements $\mathbf{O}(i)$ at pixel $i$. **O** is a vector reshaped from the initial $L \times L$ image matrix $\mathbf{O}_m$, where $L \times L = N$. The DMD is used to display the Walsh-Hadamard masks $\phi_1, \phi_2, \cdots, \phi_N$ in sequence. The $\phi_i$ with mask number $i$ ($1 \leqslant i \leqslant N$) is a $L \times L$ matrix reshaped from the $i^{th}$ row of $N$-order Walsh-Hadamard matrix $\mathbf{\Phi}$. Then, the correlation between mask and object, which is recorded by a single-pixel detector, can be mathematically described by their inner product

$$y_i = <\varphi_i, \mathbf{O}>, \tag{4}$$

where $\varphi_i$ is a $N$-length vector reshaped from $\phi_i$. The complete measurement vector is then given by

$$\mathbf{Y} = \mathbf{\Phi O}. \tag{5}$$

In experiments, the Walsh-Hadamard matrix consists of "+1" and "-1" elements were

realized by

$$\mathbf{\Phi} = \mathbf{\Phi}^{(p)} - \mathbf{\Phi}^{(n)}, \tag{6}$$

since the DMD can only modulate amplitude of the incident light. In equation (6), $\mathbf{\Phi}^{(p)}$ is constructed by submitting all "-1" elements in $\mathbf{\Phi}$ with "0". And $\mathbf{\Phi}^{(n)}$ is acquired by $\mathbf{\Phi}^{(p)} - \mathbf{\Phi}$. Ultimately, the ghost image can be reconstructed by

$$\mathbf{X} = \mathbf{\Phi}^{-1}\mathbf{Y} = \mathbf{\Phi}^{-1}(\mathbf{\Phi}\mathbf{O}) = \mathbf{O}. \tag{7}$$

**Spatiotemporal THz waveform mapping**

Let $\boldsymbol{E}(\xi,t)$ represents the THz spatiotemporal waveform to be measured, where $\xi$ and $t$ represent the spatial and temporal coordinate, respectively. $\boldsymbol{E}(\xi,t)$ consists of $N$ time-dependent vector

$$\boldsymbol{E}(\xi,t) = \begin{vmatrix} E_1(t) \\ E_2(t) \\ \vdots \\ E_N(t) \end{vmatrix}. \tag{8}$$

The sequentially recorded signals $\boldsymbol{S}(\xi,t)$ can be written as

$$\boldsymbol{S}(\xi,t) = \begin{vmatrix} S_1(t) \\ S_2(t) \\ \vdots \\ S_N(t) \end{vmatrix} = \mathbf{\Phi}\boldsymbol{E}(\xi,t). \tag{9}$$

Once the complete measurements accomplished, $\boldsymbol{E}(\xi,t)$ can be calculated as

$$\boldsymbol{E}(\xi,t) = \mathbf{\Phi}^{-1}\boldsymbol{S}(\xi,t). \tag{10}$$

In experiments, the $i^{\text{th}}$ mask had identical $N = 64$ rows (columns) for spatiotemporal waveform mapping in vertical (horizontal) direction and every row (column) of the mask was the $i^{\text{th}}$ row of $\mathbf{\Phi}$. The spatial and temporal resolution were 13 µm and 33 fs, respectively, for both Figs. 1c and 1d. And the wavefront was then regarded as a linear fit of the peak time delay in $E_i(t)$ to the corresponding spatial coordinate $\xi$. (see Supplementary Fig. S5)

**Estimation on SNR and potential frame rate of ghost image**

The SNRs of ghost images via Hadamard multiplexing (Fig. 1f) were estimated by the equation (See Supplementary section 4 for detailed mathematical derivation):

$$\text{SNR}_\text{H} = \sqrt{\frac{k}{2N[\gamma_\text{d}^2 + (\gamma_\text{s}/2)^2]}}, \qquad (11)$$

where $\gamma_\text{d}$ denotes the ratio of dark noise to THz peak, $\gamma_\text{s}$ denotes the ratio of peak RMSE (root mean square error) to THz peak, $N$ denotes the pixel number of ghost image, and $k$ denotes the number of measurements for each mask. In our experimental setup, $\gamma_\text{d}$ was measured as $1 \times 10^{-3}$, and $\gamma_\text{s}$ ~$7 \times 10^{-3}$ as indicated in Fig. 1f. Each mask value was averaged by $k$ = 15. With these parameters, the SNR$_\text{H}$ was estimated as 11.76 in the case of $N = 64 \times 64$, according to Equation (1).

The frame rate equals $\text{FPS} = 1/(2Nr_\text{c}t_\text{mask})$, where $t_\text{mask}$ represents the acquisition time for each mask and $r_\text{c}$ represents the compressive ratio. And $t_\text{mask}$ can be expressed as $t_\text{mask} = 2N(\gamma_\text{d0}^2 + \gamma_\text{s0}^2/4)t_0 \text{SNR}_\text{H}^2$, where $t_0$ denotes the pulse period, $\gamma_\text{d0}$ and $\gamma_\text{s0}$ denote the ratio of dark noise to single THz peak and pulse fluctuation ratio within the "integral time" of $t_0$, respectively. The frame rate can be derived as (See Supplementary section 4 for detailed mathematical derivation):

$$\text{FPS} = \left[N^2 r_\text{c} \text{SNR}_\text{H}^2 (4\gamma_\text{d0}^2 + \gamma_\text{s0}^2) t_0\right]^{-1}. \qquad (9)$$

As for an 80 MHz oscillator-driven GHOSTEAM system, there are reasonable values of $t_0$ = 12.5 ns, $\gamma_\text{d0}$ = 0.49, $\gamma_\text{s0}$ = 3.8 × 10$^{-3}$ (see Supplementary section 4). With these parameters, the $t_\text{mask}$ was calculated as 150 μs in the case of $N = 16 \times 16$ and SNR$_\text{H}$ = 10. The corresponding potential frame rate for full measurements was estimated as ~13 frame/s. And 50% compressive Hadamard reconstruction usually provides >95% image fidelity[2, 5], hence we estimate the potential frame rate can be further up to video rate of ~26 frames/s.

**Fs-laser manufacture**

The silica sample with three air/silica interfaces for topography was manufactured by femtosecond laser ablation method. In this method, a Ti-sapphire laser beam (800-nm central wavelength, 30-fs pulse width, 1-kHz repetition rate and 100-mW average power) was focused onto the silica substrate (1-mm thickness) using an objective lens (10×, numerical aperture NA = 0.25). The silica substrate travelled along multi-circular trajectories at the speed of 100 μm/s (original radius $r_0$ was 10 μm, interval between two adjacent circular trajectories $\Delta r$ was 10 μm, travel number $n_r$ was 10, and the radius of the inner circular was $r = r_0 + \Delta r(n_r - 1) = 100$ μm). The laser ablation depth under the above parameters was about 50 μm, the same ablation process was operated again to obtain the ablation depth of 100 μm, while the silica substrate was moved 50 μm up from the laser spot. Using the same laser fabrication method, the outer circular was produced at $R_0 = 250$ μm, $\Delta R = 10$ μm, $n_r = 15$, $R = R_0 + \Delta R(n_r - 1) = 400$ μm. And repeat the process to obtain deeper interface.

**Denoising and simulating near-field ghost topography**

The results of near-field ghost topography (Figs. 4c-4e) were denoised using the Matlab's tool box of Stationary Wavelet Transform Denoising 2-D. Five-level haar wavelet was used to decompose the images. The selected threshold method was penalize low soft thresholding.

The electromagnetic field distributions as the THz pulse propagates in the near-field (Figs. 4f-4h) were simulated using Wave Optics module of the commercial software COMSOL Multiphysics. Some certain THz wave acquired in experiment was set as the incident source and the time-dependent solver was used to resolve the electromagnetic field distributions at discrete times (time step is 33 fs). The study domain is a rectangle which is divided into silica ($n_{silica}$ = 1.97) and air ($n_{air}$ = 1), as illustrated in Fig. 4a.

## Acknowledgements

This work was supported by the National Basic Research Program of China (No. 2015CB755405), National Natural Science Foundation of China (NSAF) (Nos. U1730246, 61427814, 11604316, 11704358), the Foundation of President of China Academy of Engineering Physics (No. 201501033), and Science Challenge Project (Grant No. TZ 2018003).

## Competing financial Interests statement

L.-G. Zhu is a named inventor on Chinese patent number 201310442738.9 (publication date 27.05.2015, filing date 26.09.2013). L.-G. Zhu, J. Li, L.-H. Du, Z.-H. Zhai, and Z.-R. Li are named inventors on Chinese patent number 201410815227.1 (publication date 25.03.2015, filing date 24.12.2014) which is related to THz wave ghost microscope.

## Author Contributions

L.-G.Z., S.-C.C., Z.F., J.L. and W.T. conceived the project. S.-C.C., J.L., and Z.-H.Z. designed and performed the imaging experiments. Z.F. and W.T. designed and realized the STE-related module. J.W.C., Y.C.M., K.H. and H.F.D. fabricated the STE samples, 3D silica structure and metal patterns. Z.-R.L., C.-W.Q., and X.-C.Z. helped to design the concept. All authors discussed the results and reviewed the manuscript.